\documentclass[aps,prb,twocolumn,amsmath,amssymb,superscriptaddress,floatfix]{revtex4}

\usepackage{ulem} 

\usepackage{graphicx}
\usepackage{bm}
\usepackage[usenames]{color}
\bibstyle{apsrev.bib}

\newcommand{\be}{\begin{equation}}
\newcommand{\ee}{\end{equation}}
\newcommand{\beqn}{\begin{eqnarray}}
\newcommand{\eeqn}{\end{eqnarray}}

\def\opsimeq{\mathop{\simeq}}

\begin{document}

\title{ Locally self-similar phase diagram of the disordered Potts model on the hierarchical lattice}

\author{J-Ch. Angl\`es d'Auriac}
\affiliation{Institut N\'eel-MCBT
 CNRS 
\thanks{U.P.R. 5001 du CNRS, Laboratoire conventionn\'e
avec l'Universit\'e Joseph Fourier}, B. P. 166, F-38042 Grenoble,
France}
\author{Ferenc Igl\'oi}
\email{igloi.ferenc@wigner.mta.hu}
\affiliation{Wigner Research Centre, Institute for Solid State Physics and Optics,
H-1525 Budapest, P.O.Box 49, Hungary}
\affiliation{Institute of Theoretical Physics,
Szeged University, H-6720 Szeged, Hungary}
\date{\today}

\begin{abstract}
We study the critical behavior of the random $q$-state Potts model in the large-$q$ limit
on the diamond hierarchical lattice with an effective dimensionality $d_{\rm eff} > 2$. By varying the
temperature and the strength of the frustration the system has a phase transition line between
the paramagnetic and the ferromagnetic phases which is controlled by
four different fixed points. According to our renormalization group study the phase-boundary
in the vicinity of the multicritical point is self-similar, it is well represented by a logarithmic spiral.
We expect infinite number of reentrances in the thermodynamic limit, consequently
one can not define standard thermodynamic phases in this region.
\end{abstract}

\maketitle

\section{Introduction}
\label{sec:intr}
Disorder is an inevitable feature of real materials and it could result in dramatic changes in the
physical properties of the systems, in particular in the vicinity of phase-transition points.
The effect of disorder can be even more pronounced when both random ferromagnetic and antiferromagnetic
couplings are present. Then the combined effect of frustration as well as thermal and disorder fluctuations
could lead to a rich phase diagram, which contains a spin-glass (SG) phase and
different critical and multicritical points\cite{review_sg}.

Among the models studied in this context the random bond $q$-state Potts
model\cite{elderfield_sherrington,gross_kanter_sompolinsky,carmesin_binder,scheucher_reger_binder,schreider_reger,dillmann_janke_binder} plays an important role.
For $q=2$ we have the well known Edwards-Anderson model\cite{edwards_anderson} of Ising SG and for $q=3$ and $q=4$ it is used to
describe orientational and quadropular glasses\cite{binder_reger}, respectively. For large values of $q$ the model is 
suggested to be a plausible model of supercooled liquids. This relation seems to be correct in the
infinite-range model\cite{potts_liquid}, but for a system with finite-range interactions further studies are
needed to clarify this point. On the cubic lattice for $q=3$\cite{binder,lee} and later for $q=5$ and $6$\cite{banos} numerical evidence was obtained
in favor of a thermodynamic transition into a stable SG phase. For $q=10$ on the cubic lattice in the first studies
no SG phase has been observed\cite{binder,lee}. Repeating the calculation on a one-dimensional chain
with long-range power-law interactions, which is equivalent to a short-ranged
system below the upper critical dimension, evidence is obtained for a SG phase at low enough temperature\cite{andrist}.
Continuing this study further for even larger $q$-s seems to be very difficult, in particular due to the very
narrow range of temperature where the presence of the SG phase is expected.

In this paper, in order to gain some insight about the possible behavior of the large-$q$ model we use
the Migdal-Kadanoff renormalization group (RG) method\cite{migdal_kadanoff}, which provides exact results on hierarchical
lattices\cite{hier_lat}. This method has already been used for the Potts model (with finite value of $q$)
for random ferromagnetic couplings\cite{kinzel,derrida_gardner,andelman,monthus_garel}, while
for random ferro- and antiferromagnetic couplings this method is combined with the Gaussian approximation\cite{benyoussef}.
Studies of the model in the large-$q$ limit has been performed by one of the present authors
in Ref.\cite{paperI} to which we will refer as paper I. In paper I the lattice with an effective dimensionality,
$d_{\rm eff}=2$ is studied in more detail, in which case the critical properties of the model are calculated
by solving numerically a set of integral equations.

In this paper we continue these studies when the diamond hierarchical lattice has a larger branching number,
$b>2$, which corresponds to $d_{\rm eff}>2$. In this case the RG equations are more complicated, which are then
iterated numerically. For different values of $b$ we determine the phase diagram in the temperature-frustration plane,
which is found to contain a ferromagnetic and a paramagnetic phase, but there is no SG phase.
The phase transition
is shown to be controlled by a random ferromagnet (RF) fixed point when the driving 
force of the transition is the temperature, whereas it is controlled by 
a zero temperature (Z) fixed point
when the driving force is frustration. We calculate the critical exponents in the two attractive fixed points
for different effective dimensionalities. The attractive regimes of the RF and Z fixed points in the
phase transition line is separated by an unstable multicritical (MC) fixed point, the vicinity
of which is studied in more detail. Our numerical investigations predict that the phase-diagram around MC has a
singular structure.

The structure of the paper is the following. The model and the method of investigation is described in
Sec.\ref{sec:model}. Results about the phase diagram and the critical properties of the system are presented
in Sec.\ref{sec:PD}, while the vicinity of the MC point is studied in Sec.\ref{sec:MC}. Our paper is closed
by a Conclusion in the final section.

\section{Model and Migdal-Kadanoff renormalization}
\label{sec:model}
We consider the $q$-state Potts model defined by the Hamiltonian:
\begin{equation}
{\cal H}=-\sum_{\langle i,j \rangle}\frac{ J_{ij}}{\ln q} \delta_{s_i,s_j}
\label{hamilton}
\end{equation}
in terms of the Potts spin-variables $s_i=1,2,\dots,q$ at site $i$ of the lattice.
Here $\delta_{s_i,s_j}$ is the Kronecker-symbol, the summation runs over nearest-neighbor pairs and the couplings $J_{ij}$ are independent and identically distributed random numbers, which can be either positive or negative.
The $\ln q$ in the denominator insures that the transition temperature
stays finite when $q$ goes to infinity\cite{qlarge}.
In the random cluster representation of the model
the partition function ${\cal Z}$ is dominated by one diagram:
\begin{equation}
{\cal Z} \opsimeq_{q \to \infty} q^{\phi} + {\rm subleading~terms}
\label{dominant}
\end{equation}
and the free energy is given by $F=-\phi/\beta$, where $\beta=1/T$ with the convention $k_B=1$.

In this paper the Potts model is placed on the diamond hierarchical lattice,
which is constructed recursively from a single link. At each step a link is replaced by a
unit, which consists of $b$ parallel branches, each branch
containing two bonds in series.
At generation $n$, the length $L_n$ measured by the number of bonds between the two extreme sites
$A$ and $B$ is $L_n=2^n$, and the total number of bonds is 
\begin{eqnarray}
B_n=(2b)^n = L_n^{d_{\rm eff}(b)} \ \ \ {\rm \ \ with \  \ } 
d_{\rm eff}(b)= \frac{ \ln (2b)}{\ln 2}
\label{bn}
\end{eqnarray}
where $d_{\rm eff}(b)$ is an effective dimensionality.

In the following we consider fixed-spin boundary conditions: when the two extreme sites $A$ and $B$
are in the same (different) state the partition function is denoted by ${\cal Z}^{1,1}_n$ (${\cal Z}^{1,2}_n$).
Their ratio is given by:
\begin{equation}
\frac{{\cal Z}^{1,1}_n}{{\cal Z}^{1,2}_n}=q^{I_n}\;,
\label{x_n}
\end{equation}
where $I_n=\beta F^{\rm inter}_n$ and $F^{\rm inter}_n=F^{1,2}_n-F^{1,1}_n$ is the interface free energy.
The scaled interface free energy $I_n$ satisfies the recursion equation~\cite{kinzel,andelman}
\begin{equation}
q^{I_{n+1}}=\prod_{i=1}^b \left[ \frac{q^{I_n^{(i_1)}+I_n^{(i_2)}}+(q-1)}{q^{I_n^{(i_1)}+I_n^{(i_2)}}+(q-2)} \right]
\label{x_recursion}
\end{equation}
which for large-$q$ becomes $q$-independent (see paper I):
\begin{equation}
I_{n+1}=\sum_{i=1}^b \Phi\left[I_{n}^{(i_1)},I_{n}^{(i_2)}\right]\;.
\label{I_recursion}
\end{equation}
Here the auxiliary function is:
\begin{equation}
\Phi\left[I^{(1)},I^{(2)}\right]=  \max \left(I^{(1)}+I^{(2)},1\right)  -
\max \left(I^{(1)},I^{(2)},1\right) \;.
\label{I_recursion1}
\end{equation} 
The initial condition is given by
\begin{equation}
I_0^{(i)}=\beta J_i
\end{equation}
where $J_i$ is the value of the $i$th coupling.

\section{Phase diagram and critical properties}
\label{sec:PD}

\subsection{Numerical pool method}
\label{sec:pool}

The phase diagram of the random system is studied numerically for different numbers of the
branching number $b>2$, which corresponds to an effective dimension $d_{\rm eff}>2$.
The original distribution of the couplings is given in a box-like form:
\begin{equation}
{\cal P}(J)=
\begin{cases}
1 & \text{if $\frac{p}{1-p}<J<\frac{1}{1-p}$},\\
0& \text{otherwise}.
\end{cases}
\label{distr_SG}
\end{equation}
with $p \le 1$ as already used in Paper I for $b=2$.
For $p>0$ all couplings are random ferromagnetic and in the limit $p \to 1$ we have the pure system.
For $p<0$ there are also negative bonds, their fraction is increasing with decreasing $p$.
In the numerical calculations
we have used the so-called pool method. Starting with $N$ random variables taken from the original distribution in
Eq.~(\ref{distr_SG}),  we generate a new set of $N$ variables through renormalization at a fixed temperature, $T$,
using Eqs.~(\ref{I_recursion}) and (\ref{I_recursion1}). These are the elements of the pool at the first generation, which are then used as input for 
the next renormalization step.
We check the properties of the pool at each renormalization by calculating the distribution of the scaled interface free energy, its average
and its variance. In practice we have used a pool of $N=5 \times 10^6$ elements and we went up to $n \sim 70-80$ iterations.
We mention that in order to study the sensitivity of the results on the form
of the initial distribution we have also 
used Gaussian- and symmetric bimodal
distributions instead of Eq.(\ref{distr_SG}), but the structure of the phase diagram and the
properties of the fixed points remain the same.

\subsection{Phases}
\label{sec:phases}
At a given point of the phase diagram, $(p,T)$, the renormalized parameters display two different behaviors, which are governed by two trivial fixed points.

i) In the paramagnetic phase the scaled interface free energy renormalizes to zero, thus its average is
$\overline{I}=0$ and its variance is $\Delta I=0$.

ii) In the ferromagnetic phase the scaled interface free energy goes to infinity. Its average behaves at a
length, $L$ as $\overline{I} \sim L^{d_s}$, with $d_s=d_{\rm eff}-1$ being the effective dimension of the interface.
Also the variance of $I$ is divergent: $\Delta I \sim L^{\theta}$, with a droplet exponent $\theta>0$. The length $L$ has been 
defined by Eq.\ref{bn}.

We note that in the
large-$q$ limit there is no spin-glass phase, even at zero-temperature, contrary to the known results for $q=2$ and $q=3$.\cite{binder1,huse}.


\begin{figure}[h!]
\begin{center}
\includegraphics[width=8.cm,angle=0]{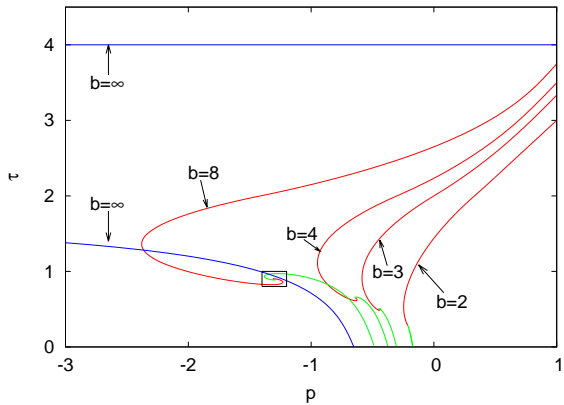}
\caption{Phase-diagrams for different branching numbers: $b=2,~3,~4,~8$
and in the $b \to \infty$ limit. The system is in the paramagnetic phase for high temperature (large $T$ and $\tau$) and/or
for large frustration (small $p$), otherwise it is in the ferromagnetic phase. The phase transition lines are
indicated by red and green, along which the transition is controlled by the RF and the Z fixed points,
respectively. The red and green lines meet at the MC points where there are several re-entrances for $b>2$.}
\label{Fig1}
\end{center}
\end{figure}


The two phases
are separated by a phase transition line $T_c(p)$ which can be calculated accurately for a given pool and its true value can be obtained by averaging over
different pools and taking the $N \to \infty$ limit. The phase diagrams are shown in Fig.~\ref{Fig1} for different values
of the branching number in the plane $\tau=(1+p)T/\overline{J}$ versus $p$. (In the vertical axis we used a nonsingular scaling combination, since the average coupling:
$\overline{J} = \frac{p+1}{2}$ is zero at $p=-1$.)
For the case  $b\rightarrow \infty$, the exact expressions of the transition
lines can be obtained and these are given in Appendix. Interestingly
the phase diagram shows several reentrances in the regime $p<0$. 
(see also Fig.~\ref{Fig5}).
This behavior is different from that observed
at $b=2$ (in which case only one reentrance is present) and we are going to study its properties
in more detail in Sec. \ref{sec:MC}.

\subsection{Fixed points}
\label{sec:fpoints}

\begin{figure}
\begin{center}
\includegraphics[width=8.cm,angle=0]{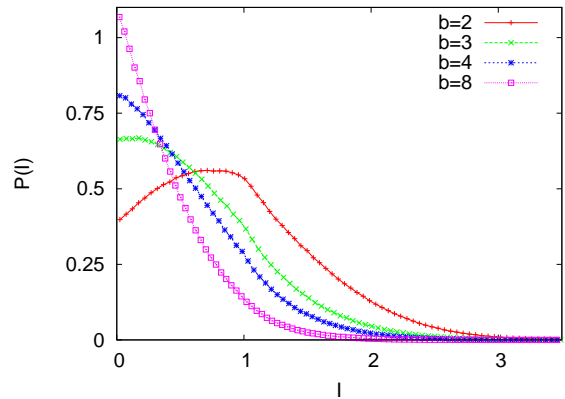}
\end{center}
\caption{(Color online) Probability distribution of the scaled interface free energy at the RF fixed point for
$b=2$, $b=3$, $b=4$ and $b=8$. At $I=0$ there is
an extra delta peak with respective strength 0.12808, 0.25052, 0.31889, and 0.43609. }
\label{Fig2}
\end{figure}

For all values of $b$ the phase-transition line is found to be controlled by four different fixed points.

i) The fixed point of the pure system (P) is located at $p=1$ and $\tau=2(2b-1)/b$ and it describes a
first-order transition. This fixed point is unstable for any
small amount of disorder, see paper I.

ii) Along the transition line for $p<1$ between P and the multicritical point MC (indicated
by red line in Fig. \ref{Fig1}) the transition
is controlled by the RF fixed point. Here $\overline{I}={\cal O}(1)$
and $\Delta I={\cal O}(1)$ and the distribution
$P(I)$
at the fixed point is given in Fig.\ref{Fig2} for $b=2,~3,~4$, and $8$. At this fixed point
the distribution contains no negative $I$ values, but there is a $\delta$-peak at $I=0$, the strength
of which is increasing with $b$, see the caption of Fig.\ref{Fig2}.
In the vicinity of the transition line at a distance $\Delta T=T_c-T$ in the ferromagnetic
phase the average value and the variance of $I$ behave as:
\beqn
\overline{I}(\Delta T,L)&=&\left[\frac{L}{\xi_{\rm av}(\Delta T)}\right]^{d_s}+\dots,\cr
\Delta I(\Delta T,L)&=&\left[\frac{L}{\xi_{\rm var}(\Delta T)}\right]^{\theta}+\dots\;.
\label{IL}
\eeqn
Here the correlation lengths are divergent as: $\xi_{\rm av}(\Delta T) \sim (\Delta T)^{-\nu_{\rm av}}$ and
$\xi_{\rm var}(\Delta T) \sim (\Delta T)^{-\nu_{\rm var}}$, respectively.


iii) The other part of the transition line between MC and Z (indicated
by green line in Fig. \ref{Fig1}) is controlled by the
Z fixed point, which is located at $T=\tau=0$. Here the interface free-energy grows with the size as:
$I(L)=L^{\theta_Z} u_I$, where the $u_I$ are ${\cal O}(1)$ random numbers and the droplet
exponent is $\theta_Z>0$. Here we use the scaled variable, $i_n \equiv I_n/\overline{I}_n$, which
obeys the renormalization group equation from Eq.(\ref{I_recursion}):
\be
\frac{I_{n+1}}{\overline{I}_n}=i_{n+1} \alpha_{n+1}==\sum_{i=1}^b \phi\left[i_{n}^{(i_1)},i_{n}^{(i_2)}\right]\;.
\label{i_recursion}
\ee
with $\alpha_{n+1}=\overline{I}_{n+1}/\overline{I}_n$ and
\begin{equation}
\phi\left[i^{(1)},i^{(2)}\right]=  \max \left(i^{(1)}+i^{(2)},0\right)  -
\max \left(i^{(1)},i^{(2)},0\right) \;.
\label{i_recursion1}
\end{equation}
The fixed-point value of the ratio: $\alpha^Z$ is related to the droplet exponent as: $\theta_Z=\log \alpha^Z/\log 2$. 

\begin{figure}
\begin{center}
\includegraphics[width=8.cm,angle=0]{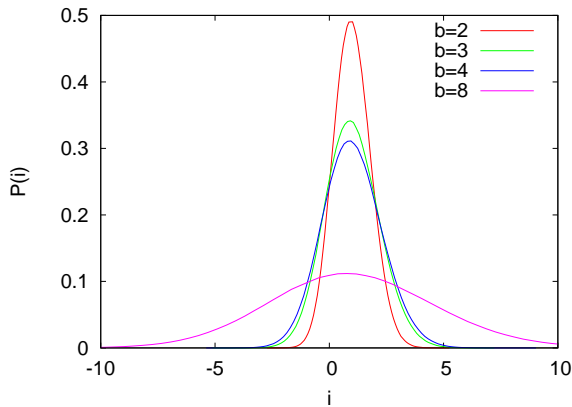}
\end{center}
\caption{(Color online) Probability distribution of the scaled interface free energy at the Z fixed point. At $i=0$ there is a delta peak with strength
$0.1728 \,10^{-3}$ for $b=2$, $0.4928  10^{-4}$ for $b=3$, 
and $0.9400 10^{-5}$ for $b=4$. For $b=8$ and with a pool of size $5 \times 10^6$
we have found no zero value.}
\label{Fig3}
\end{figure}

The distribution of the scaled variable: $i$
is shown in Fig.\ref{Fig3} for various values of $b$.
In the vicinity of the transition line at a distance $\Delta T=T_c^{'}-T^{'}$
in the ferromagnetic phase the average value and the variance of the interface free energy scales as:
\beqn
\overline{I}(\Delta T,L)=\frac{L^{d_s}}{[\xi_{\rm av}(\Delta T)]^{d_s-\theta_{\rm Z}}}+\dots\;,\cr
\Delta I(\Delta T,L)=\frac{L^{\theta}}{[\xi_{\rm var}(\Delta T)]^{\theta-\theta_{\rm Z}}}+\dots\;,
\label{IZ}
\eeqn
respectively. Here the correlation lengths, $\xi_{\rm av}(\Delta T)$ and $\xi_{\rm var}(\Delta T)$ are
divergent at $\Delta T=0$, with the critical exponents: $\nu_{\rm av}^{Z}$ and $\nu_{\rm var}^{Z}$, respectively.


\begin{figure}
\begin{center}
\includegraphics[width=6.cm,angle=270]{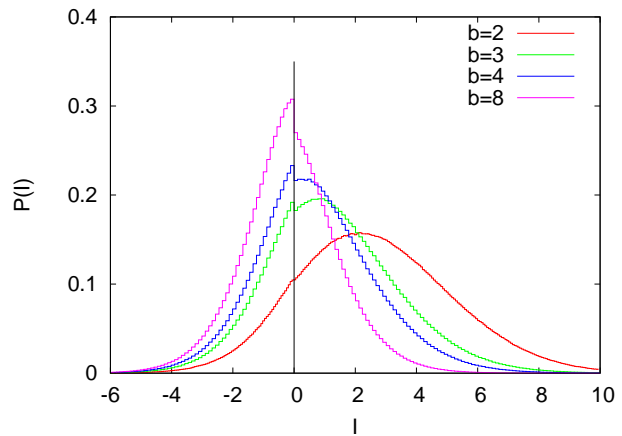}
\end{center}
\caption{(Color online) Probability distribution of the scaled interface free energy at the MC fixed point
for $b=2,~3,~4$ and $8$. For $b>2$ there is a discontinuity at $I=0$ (a {\it step} 
representation of the probability distribution has been used to stress 
this feature). In addition there is a delta peak at $I=0$ with
respective weight 0.0057, 0.012854, 0.014713 and 0.016626 for $b=3$, $b=4$ and
$b=8$.}
\label{Fig4}
\end{figure}

iv) The attractive regions of the RF and Z fixed points are separated by the
unstable MC fixed point. Here the interface free energy behaves as: $\overline{I}={\cal O}(1)$
and $\Delta I={\cal O}(1)$, similarly to the RF fixed point. The distribution, $P(I)$ at the MC
fixed point is shown in Fig.\ref{Fig4} for various values of $b$. This distribution has a finite jump at $I=0$
for $b>2$, which has interesting consequences about the structure of the phase diagram
at MC, what we are going to analyze in Sec.\ref{sec:MC}.


\subsection{Critical exponents}
\label{sec:exponents}
The critical exponents for $b>2$ are calculated numerically by the pool method.
In the calculation  of the RF exponents we have first fixed the parameter of
the disorder distribution, $p$, then calculated the location of the critical point: $T_c(p)$. For a given
pool $T_c(p)$ has been calculated accurately, with a precision of $10^{-12}$. Then estimates for the critical exponents are calculated from the relations in Eqs.(\ref{IL}). At the Z fixed point the temperature is $T=0$
therefore we have to calculate its position: $p_Z$. Then the critical exponents
are calculated from (\ref{IZ}). Repeating the
procedures for different parameters and for different pools we have obtained estimates of the set of critical exponents
with a relative precision of about $10^{-3}$. We have observed that the correlation length critical
exponents calculated from the average and from the variance are identical for a given fixed point, i.e.
$\nu_{\rm av}=\nu_{\rm var}=\nu$ and similarly $\nu_{\rm av}^Z=\nu_{\rm var}^Z=\nu^Z$. This relation is found to be
exact before for $b=2$ (see paper I) and now it is demonstrated numerically for other values of the branching number.
We note that for finite value of $q$ the two correlation length exponents are generally different,
at least on the same diamond hierarchical lattice\cite{monthus_garel}. The set of numerically calculated critical exponents are
collected in Table \ref{table:1} for different values of the branching number. For completeness we also present here
the numerically exact results obtained for $b=2$ in paper I.

\begin{table}[h]
\caption{Critical exponents of the disordered Potts model in the large-$q$ limit
on the diamond hierarchical lattice.
Results for branching number $b=2$ are from paper I.\label{table:1}}
 \begin{tabular}{|c|c|c|c|c|c|c|}  \hline
   $b$& $2$ & $3$ & $4$ & $8$ & $16$ & $32$ \\ \hline
$\theta$& $0.299$ & $0.563$ & $0.760$ & $1.245$ & $1.736$ & $2.233$ \\\hline
$\nu$&$1.307$  & $1.051$& $0.949$ & $0.808$ & $0.730$ & $0.684$ \\ \hline
$\theta_Z$& $0.146$ & $0.325$ & $0.478$& $0.894$ & $1.348$ & $1.822$ \\
$\nu^Z$&$1.729$  & $1.258$ & $1.051$& $0.738$ & $0.555$ & $0.440$ \\ \hline
  \end{tabular}
  \end{table}

\section{Multicritical point}
\label{sec:MC}
\begin{figure}[!ht]
\begin{center}
\includegraphics[width=8.cm,angle=0]{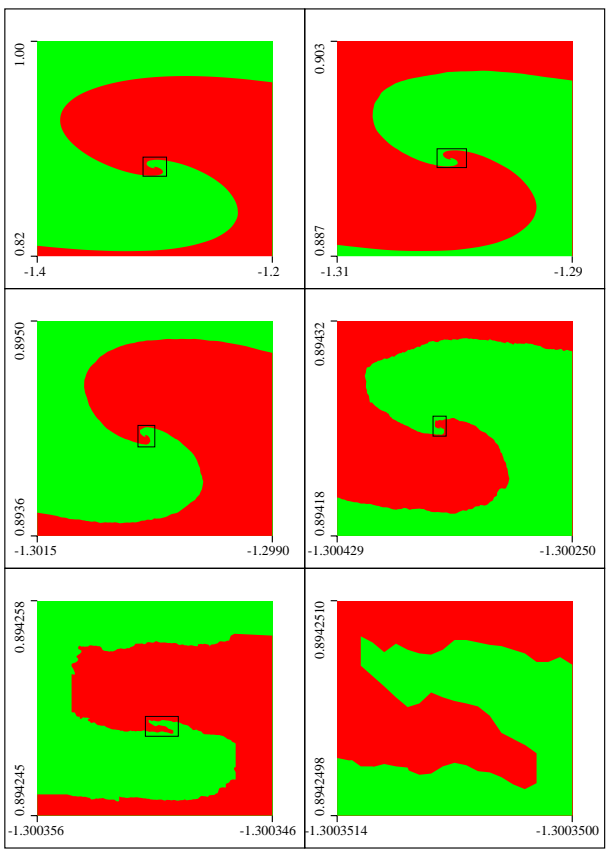}
\caption{Phase-diagram in the vicinity of the MP point: the
framed regions are enlarged in the next figure.
Note the (self-)similarity of the phase boundary in different scales.
The green (bright) region represents the paramagnetic phase, the red (dark) region is the ferromagnetic phase.}
\label{Fig5}
\end{center}

\end{figure}

The phase-diagram in the vicinity of the MC point is different for $b=2$ and for $b>2$, see
in Fig.\ref{Fig1}. In the former case there is just one reentrance of the phase-boundary and at the
MC fixed point there is traditional multicritical behavior, see in paper I. On the contrary
for $b>2$ the phase diagram around MC has more than one reentrance. We have explored the phase
diagram around MC by using larger and larger magnifications and the results for $b=8$ are
presented in Fig.\ref{Fig5}. Interestingly the phase-diagram shows a self-similar structure: the reentrances
are repeated in different scales. Let us denote the location of the MC point by: $\tau^*$
and $p^*$ and then use relative positions of the phase-boundary, which are given by $\Delta p=p_c-p^*$ and
$\Delta \tau=\tau_c-\tau^*$. According to the results in Fig.\ref{Fig5} by enlarging the scale by roughly a factor
of ten the phase-diagram has approximately the same reentrant structure. With the pool method, having a
large, but finite number of realizations we could go up to a magnification of $\rho_0 \approx 2 \times 10^{-7}$ in the
phase diagram, up to which scale one could observe at least seven similar reentrant structures.


\begin{figure}[!ht]
\begin{center}
\includegraphics[width=6.cm,angle=0]{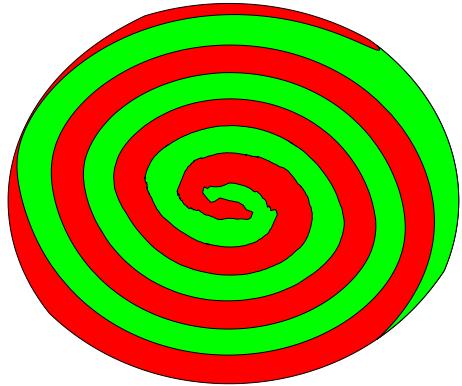}
\caption{Phase-diagram in the vicinity of the MP point in logarithmic polar coordinates, see text.}
\label{Fig6}
\end{center}

\end{figure}

It is instructive to use polar coordinates, $\rho$ and $\phi$, and plot the phase-diagram in a semi-logarithmic
scale as a function of $\ln(\rho/\rho_0)$ and $\phi$, which is presented in Fig.\ref{Fig6}. In this figure
the paramagnetic and the ferromagnetic phases are represented by two almost perfect spirals, which means that the phase-boundary around the MC point is approximately given by a logarithmic spiral.
We note that for other values of the branching number ($b=3$ and $b=4$) we have observed a similar structure of the phase diagram around the MC point.


We expect that the size of the elementary magnification step, $\rho_0$, is decreasing when the
size of the pool is increasing and in the thermodynamic limit the
magnification step goes to zero.
 In this case the phase diagram at the MC point becomes
singular. Passing through the MC point the phase diagram has infinite number of reentrances and sufficiently
close to MC one can not define a stable thermodynamic phase, since its stability range measured either
in temperature or in concentration of couplings goes to zero.

\section{Conclusion}
\label{sec:Concl}
In this paper we considered the $q$-state Potts model for large $q$ with randomly mixed ferro- and antiferromagnetic
couplings and studied its critical properties on the diamond hierarchical lattice with an effective dimensionality,
$d_{\rm eff} >2$. By using the Migdal-Kadanoff
renormalization scheme, which is exact for this lattice, the phase diagram is explored numerically using
the pool method. As for $d_{\rm eff} = 2$ the phase diagram contains a ferromagnetic and a paramagnetic phase, but
there is no spin glass phase, even at zero temperature. The phase transition in the random system is controlled
either by a random ferromagnet fixed point, in which the driving force is the temperature, or by a zero temperature
fixed point, in which the transition is due to frustration. The two regimes are separated by a multicritical
point in the vicinity of which the system shows unusual properties.

Both in the RF and the Z fixed points the transition is continuous and we have calculated numerically the
critical exponents, which are collected in Table \ref{table:1}. The correlation length exponents calculated
from the scaling of the average or the variance are found to be equal. At the MC point the distribution of the
reduced interface free energy is a discontinuous function, having a finite jump at $I=0$. At the same time
the phase-diagram close to MC has a self-similar, logarithmic spiral-like structure. Consequently the
phase boundary itself has a singularity and in the
traditional sense one can not define a stable thermodynamic phase here.

The studies presented in this paper can be extended in different directions. First, it would be interesting to study by
analytical methods the behavior of the system in the vicinity of the MC point for $b>2$. i) To prove or disprove,
that the elementary magnification step at the MC point, $\rho_0$, approaches zero
in the thermodynamic limit. ii) To show rigorously, that the distribution function, $P(I)$, in
the MC point is strictly discontinuous, as shown in Fig.\ref{Fig4}. iii) Also would be interesting to clarify the
possible relation between these two expected properties. 

Another extention of the present study is to investigate, if the
singular behavior at the MC point could exist in another
systems, too. As far as the random Potts model on hypercubic lattices is concerned it has the same fixed
point structure for finite value of $q$, thus generally there is an MC point in these models\cite{huse,picco}.
Probably in sufficiently large dimension ($d=3$ could be enough), and for large-$q$, the MC singularity could
exist. Numerically, however, it would be very difficult to explore the details of the process, as already seen
in the problem of detection of the SG phase\cite{banos,andrist}. A discontinuous distribution of some observables at this point,
such as the reduced interface free energy, could be the sign of such singularity.

\begin{acknowledgments}
The authors thank to each other Institutes for facilitating visits. F.I. thanks to C. Monthus and L. Turban for collaboration
in related problems. He is indebted to A. N. Berker for useful discussions and to M. A. Moore for a helpful correspondence.
This work has been supported by the Hungarian National Research Fund under grant
No OTKA K75324 and K77629.
\end{acknowledgments}

\section*{Appendix: The phase-diagram for $b \to \infty$}
\label{appendix}
The phase-diagram for $b\rightarrow \infty$ can be analytically treated,
here we consider the initial distribution in Eq.(\ref{distr_SG}).
First we note that in this limit in the recursion equation Eq.(\ref{I_recursion})
there is a sum of a large number of random variables and the renormalized value of
the interface free energy depends on the sign of the average value of the auxiliary
function: $\left<\Phi(I_0^{(1)},I_0^{(2)})\right> \equiv \left<\Phi\right>$. Here we can differentiate between
three cases. 

i) If $\left<\Phi\right> > 0$
then all the renormalized parameters are $I_1 \gg 1$, consequently we are at the F fixed point
and due to the central limit theorem the droplet exponent is $\theta=d_s/2$.

ii) If $\left<\Phi\right> = 0$,
then all the $I_1$ values are equal to zero, thus
the P fixed point has been reached. 

iii) Finally, if $\left<\Phi\right> < 0$, then all the renormalized
parameters are $I_1 \ll 0$, thus in the next renormalization step we have $I_2=0$ and again the P fixed point
has been reached.

In the next step we compute the sign of $\left<\Phi\right>$ in the
different parts of the phase diagram in Fig.\ref{Fig1}. Let us denote the end-points
of the box-distribution in Eq.(\ref{distr_SG}) by $x=\frac{2p}{\tau}$ and $y=\frac{2}{\tau}$. 
In the region $y<1$ one has $\left<\Phi\right>=0$ and only for
$x<0$ and $y>1$ we
can have $\left<\Phi\right><0$. A direct calculation gives in this region
$\left<\Phi\right>$
as the ratio of a third degree polynomial in $x$ and $y$ by $(y-x)^2$ and
the numerator has two different expressions according to the
sign of $(x+y+1)$. Then expressing $x$ and $y$ with
$p$ and $\tau$, one gets that the transition lines
shown in Fig.~\ref{Fig1}. The upper boundary is a
horizontal line with $\tau=4$. The lower boundary is given by two expressions:
\begin{eqnarray*}
-3\tau^3+12\tau^2-24\tau+32+12p\tau^2-48p\tau+48p & = & 0\\
-16p^3-48p^2+24p^2\tau-\tau^3+16 & = & 0\;,
\end{eqnarray*}
which are valid for $p<\tilde{p}$ and $p>\tilde{p}$, respectively.
The two lines join at:
\begin{eqnarray*}
\tilde{p}&=&\frac{1}{3}\left(1-4\cos\theta\right)\\
\tilde{\tau}&=&\frac{8}{3}\left(1-\cos\theta\right)\\
\theta&=&\frac{\arctan\left(3\sqrt{7}\right)-\pi}{3}\;,
\end{eqnarray*}
and the extremities of the lower boundary line are at $p=-\infty$,
$\tau=2$ and $p=2\cos\left(\frac{4}{9}\pi\right)-1$, $\tau=0$.

We note that in the limit $b \to \infty$ the ferromagnetic phase between $2 < \tau < 4$
extends to all values of $p$. For large, but finite value of $b$, however, there is a paramagnetic phase
for $p \lesssim -\sqrt{b}$.

\end{document}